\documentclass[a4paper,fleqn,usenatbib]{mnras}

\usepackage[T1]{fontenc}
\usepackage{ae,aecompl}

\newcommand{\lsim}{\mathrel{\rlap{\raise -.3ex\hbox{${\scriptstyle\sim}$}}%
		  \raise .6ex\hbox{${\scriptstyle <}$}}}%
\newcommand{\gsim}{\mathrel{\rlap{\raise -.3ex\hbox{${\scriptstyle\sim}$}}%
		  \raise .6ex\hbox{${\scriptstyle >}$}}}%

\usepackage{graphicx}	
\usepackage{amsmath}	
\usepackage{amssymb}	

\title[Satellite remote sensing of light pollution sources]{Aerosol characterization using satellite remote sensing of light pollution sources at night}

\author[M. Kocifaj et al.]{
Miroslav Kocifaj,$^{1,2}$\thanks{E-mail: kocifaj@savba.sk}
Salvador Bar\'{a},$^{3}$
\\
$^{1}$Faculty of Mathematics, Physics, and Informatics, Comenius University, Mlynsk\'{a} dolina, 842 48 Bratislava, Slovakia\\
$^{2}$ICA, Slovak Academy of Sciences, D\'{u}bravsk\'{a} cesta 9, 845 03 Bratislava, Slovakia\\
$^{3}$Departamento de F\'{\i}sica Aplicada, Universidade de Santiago de Compostela, 15782 Santiago de Compostela, Galicia, Spain
}

\date{Accepted XXX. Received YYY; in original form ZZZ}

\pubyear{2020}

\begin{document}
\label{firstpage}
\pagerange{\pageref{firstpage}--\pageref{lastpage}}
\maketitle

\begin{abstract}
A demanding challenge in atmospheric research is the night-time characterization of 
aerosols using passive techniques, that is, by extracting information from scattered 
light that has not been emitted by the observer. Satellite observations of artificial
night-time lights have been used to retrieve some basic integral parameters, like the 
aerosol optical depth. However, a thorough analysis of the scattering processes allows 
one to obtain substantially more detailed information on aerosol properties. In this 
Letter we demonstrate a practicable approach for determining the aerosol particle 
size number distribution function in the air column, based on the measurement of the 
angular radiance distribution of the scattered light emitted by night-time lights of 
cities and towns, recorded from low Earth orbit. The method is self-calibrating and  
does not require the knowledge of the absolute city emissions. The input radiance 
data are readily available from several spaceborne platforms, like the VIIRS-DNB 
radiometer onboard the Suomi-NPP satellite.

\end{abstract}

\begin{keywords}
light pollution -- scattering -- radiative transfer -- atmospheric effects -- instrumentation: photometers
\end{keywords}



\section{Introduction}

Determining the aerosol properties at night-time is an essential step for a better 
understanding of aerosol dynamics, with direct applications to the study of planetary 
atmospheres, the characterization of potential candidate sites for astrophysical 
observatories, and light pollution research. Artificial night lights of cities and 
towns offer a permanent set of light beacons distributed worldwide, whose observation 
from Earth orbiting platforms has been shown to be useful to determine some basic integral 
properties of the aerosols contained in the air colum, as e.g. the aerosol optical depth 
\citep[]{ChooJeong2016, JohnsonEtAl2012, McHardyEtAl2015, WangEtAl2016, ZhangEtAl2008, 
ZhangEtAl2019}. However, night-time imagery contains additional useful information, 
among other the angular dependence of the signal produced by the artificial light 
scattered by the atmosphere, that can be measured in clear and moonless nights by 
observing the radiance received at the satellite from directions corresponding to Earth 
pixels with no light sources, preferably located in the vicinity of urban nuclei with 
sharp geographical borders. This scattered light appears in the satellite images as a 
diffuse glow surrounding the city areas, blurring the city limits, and with decreasing 
radiance for increasing distances \citep[e.g.][]{SanchezEtAl2019}. In this Letter we 
show that the specific way in which the scattered radiance varies with the nadir angle 
of the observed pixels, as seen from the satellite radiometers, can be used to retrieve 
more specific key aerosol properties like e.g. their particle size number distribution 
function, if some additional {\it a-priori} information is available. An important 
feature of this approach is that it is self-calibrating: since it is based on the 
analysis of the angular dependence of the scattered radiances normalized to the direct 
radiance received from the city, the knowledge of the absolute emissions from the urban 
area is not required for its application.

\section{The model}
\label{sec:model}

A light source seen in satellite imagery of an atmosphereless planet would be geometrically 
sharply confined. However, due to light scattering by atmospheric constituents, the 
brightness of the artificially lit surface of the Earth at night does not change abruptly 
when transitioning from bright image pixels to neighbouring pixels at the outer interface 
of a city or town. Normally the radiance of a surface decreases steeply but in a continuous 
way, as the angular distance from the city-edge increases, while quickly approaching the 
background level.

\begin{figure}
	\includegraphics[width=1\columnwidth]{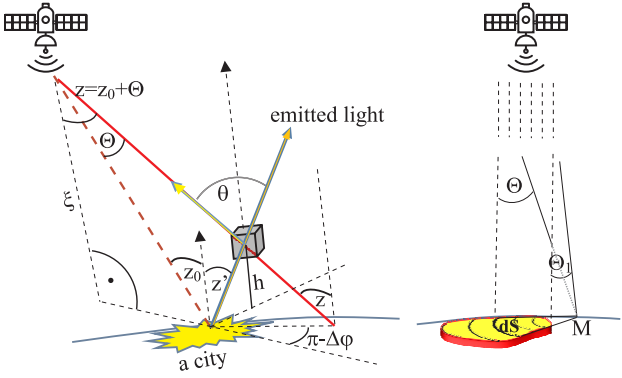}
    \caption{Scattering geometry for: (a)(left plot) remote sensing away from the satellite 
	vertical direction, (b)(right plot) satellite at near zero zenith angle from the city.
	The parameters shown in the figure are described in the main body of the text.}
	\label{fig:scatt_scheme}
\end{figure}

\begin{figure}
	\begin{minipage}[t]{0.48\columnwidth}
	\includegraphics[width=1\columnwidth]{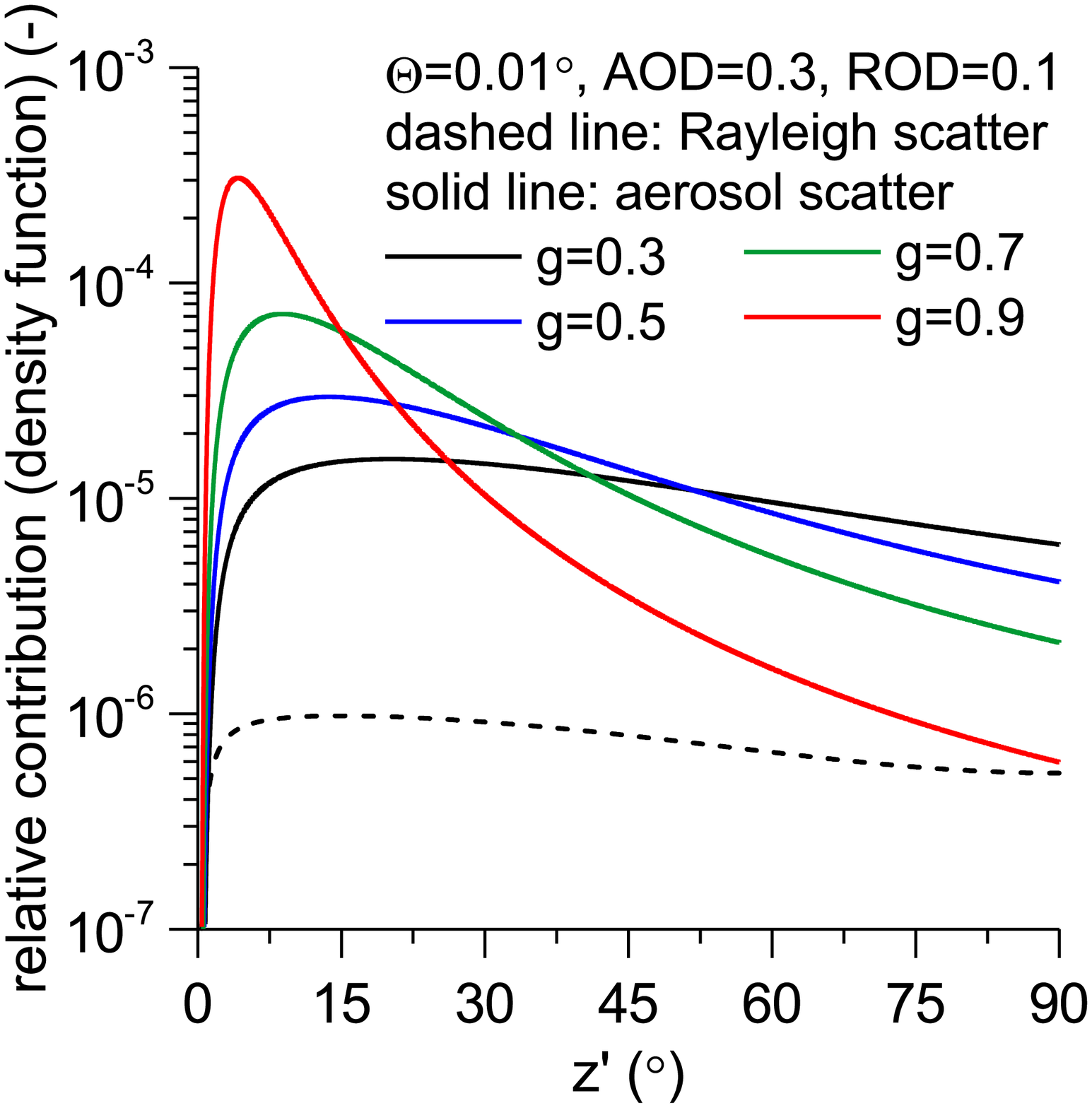}
	\end{minipage}
	\begin{minipage}[t]{0.48\columnwidth} 
	\includegraphics[width=1\columnwidth]{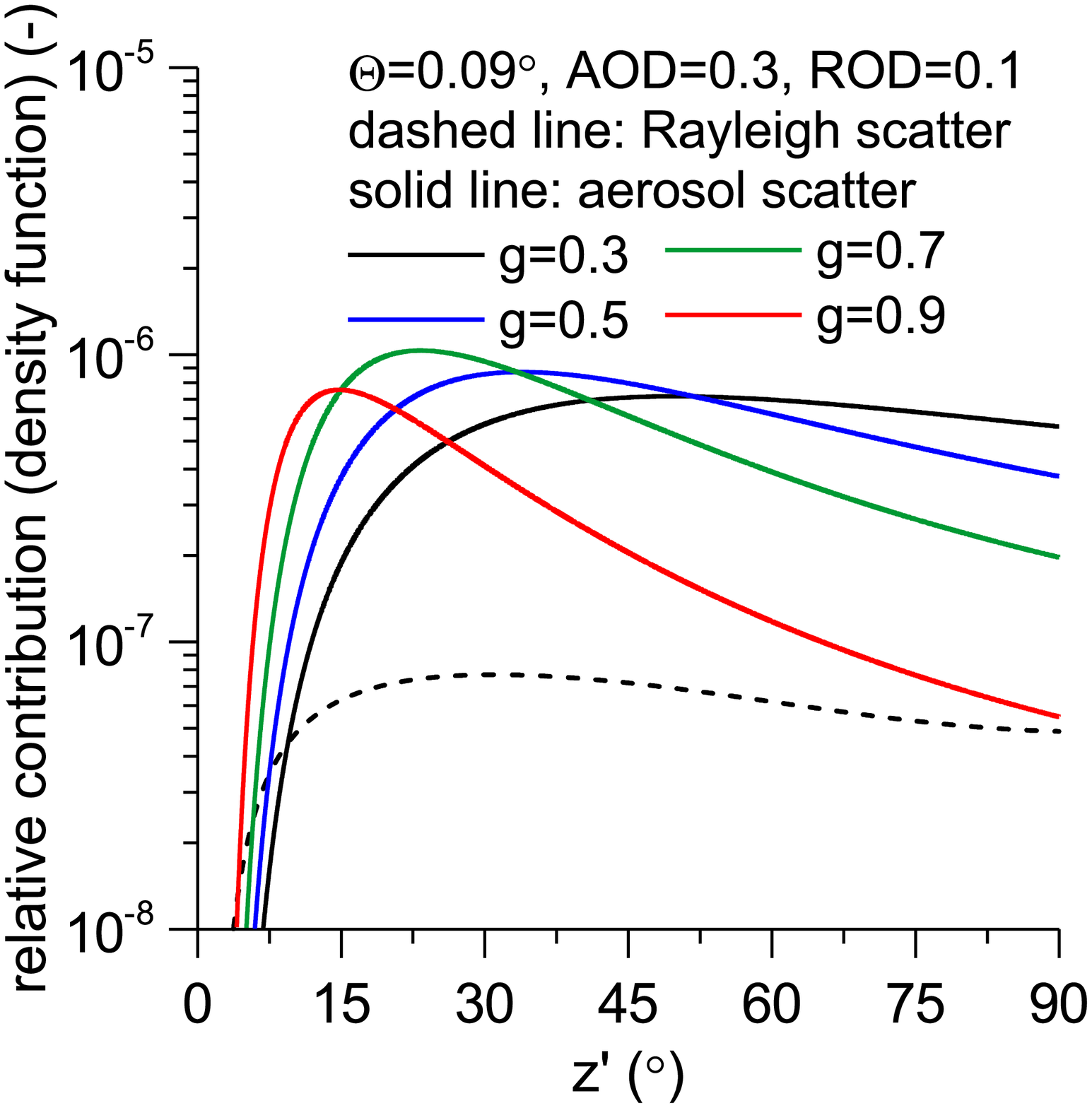}
	\end{minipage} 
    \caption{The contributions from aerosol and Rayleigh scatterings to the 
	radiance at satellite level, shown as a function of the emission angle $z'$ 
	(see Fig.~\ref{fig:scatt_scheme}a). Left plot (a) is for $\Theta=0.01^{\circ}$, 
	while the right plot (b) is for $\Theta=0.09^{\circ}$. The aerosol populations 
	under analysis differ in the asymmetry parameter, $g$, which can theoretically 
	range from -1 (entirely backscattered light) to 1 (scattered light 	directed 
	exclusively in the forward direction). Negative values of $g$ are scarcely 
	observed in nature. ROD is the Rayleigh Optical Depth, while AOD is the
	Aerosol Optical Depth.}
    \label{fig:Rayleigh_vs_Mie}
\end{figure}

The radiance of the Earth's surface, $R_{S}(z)$, reaching the satellite sensor can be 
computed as the integral of elementary optical signals produced from light scattering 
in the atmospheric volumes along the line of sight 
(see Fig.~\ref{fig:scatt_scheme}a)
\begin{eqnarray}\label{eq:R_S}
    R_{S}(z) &=& \frac{1}{\cos z} \int_{h=0}^{\xi} dI_{0}(z') \exp \left \{ 
	-\frac{\tau(h)}{\cos z} \right .
	\nonumber \\
	&& \left . -\frac{\tau(0)-\tau(h)}{\cos z'} \right \} \tilde{\omega} k_{ext}(h) 
	\frac{\cos^{2} z'}{h^{2}} \frac{p(\theta)}{4\pi} dh~,
\end{eqnarray}
where $z \approx z_{0} + \Theta$ is the zenith angle of a surface area in the city 
surroundings that has no light sources, while $z_{0}$ is the zenith angle of an 
illuminated surface element within the city or town. In Fig.~\ref{fig:scatt_scheme}a 
$\Theta$ is the angle between the direction of observation and the position of a point s
ource of light (or a surface element $dS$ of an artificially lit area). The normalized 
scattering phase function $p(\theta)/(4\pi)$ shapes with the scattering angle $\theta$ 
\begin{eqnarray}\label{eq:ScattAngle}
    \cos \theta &=& \cos z ~\cos z' + \sin z ~\sin z' ~\cos (\Delta \varphi)~.
\end{eqnarray}
The angles $z$, $z'$ and $\Delta \varphi$ are defined in Fig.~\ref{fig:scatt_scheme}a. 
In Eq. \ref{eq:R_S} the product of the single scattering albedo, $\tilde{\omega}$, and the 
volume extinction coefficient, $k_{ext}$, is known as the atmospheric volume scattering 
coefficient \citep[see e.g.][]{Kokhanovsky1998}. The optical thickness $\tau(h)$ is 
calculated for an atmospheric column that extends from altitude $h$ to the satellite 
level $\xi$, i.e. $\tau(h) = \int_{h}^{\xi} k_{ext}(h') dh'$. Finally,  $dI_{0}(z')$ is 
the radiant intensity (measured in $W sr^{-1}$) emitted by an elementary area dS of the 
city, that can be expressed as 
\begin{eqnarray}\label{eq:omgsc}
    dI_{0}(z') &=& \tilde{R}_{0}(z') dS ,
\end{eqnarray}
where $ \tilde{R}_{0}(z')$ has the units of radiance ($W m^{-2} sr^{-1}$). The radiant 
intensity emitted by a city normally depends on the emission angle $z'$. The 
collective effect of differently oriented and sloped surfaces, however, is that the 
intensity usually varies slowly with  $z'$. This is consistent with the models 
currently in use \citep[e.g. Fig. 5 in][]{KollathEtAl2014}. The angular dependences 
of the remaining terms in the integrand in Eq. \ref{eq:R_S} are much more pronounced, e.g. 
$p(\theta)$ shows a dramatic decrease rate by one or more orders of magnitude across  
its definition domain, with a typical strong forward lobe and a weak side scattering. After 
a considerable manipulation of mathematical expressions (not shown here) we obtain 

\begin{eqnarray}\label{eq:RS_radiance}
    R_{S}(\Theta) &=& \frac{\tilde{R}_{0} e^{-\tau(0)}}{(\xi \sin \Theta)^{2}} dS \left (
	\frac{\lambda}{2\pi} \right )^{2}
	\nonumber \\
	&& \int_{0}^{\xi} \left [ \int_{0}^{\infty} S_{11}(a,z') f(a,h) da \right ]
	\sin^{2}(z') dh~,
\end{eqnarray} 
for the angular dependence of the scattered radiance detected by the satellite at 
wavelength $\lambda$. In this equation it is assumed that the satellite passes almost 
vertically over the city (i.e. at near zero zenith angle), and that the radiant 
intensity emitted by the city is slowly dependent on $z'$, having an average value 
$\tilde{R}_{0}dS$. Due to atmospheric extinction the radiance of the bright pixels is 
reduced to $\tilde{R}_{0\xi} = e^{-\tau(0)} \tilde{R}_{0}$ when measured at the satellite 
level. The approximations used in deriving the above formula also comprise i) 
$\theta \approx z'+\Theta \approx z'$, ii) $z=\Theta$, iii) the contribution of Rayleigh 
scattering to the radiance recorded at satellite level for $\Theta \lsim 0.09^{\circ}$ 
is much smaller than the one from aerosols (see Fig.~\ref{fig:Rayleigh_vs_Mie}), and iv) 
the Mie theory for spherical homogeneous particles of radii $a$ controls the far-field 
scattering amplitude through the first element of the scattering matrix, $S_{11}$,  which 
is also known as the scattering function. The scattered signal from an ensemble of 
independent particles is proportional to the product of $S_{11}(a,z')$ and $f(a,h)$, where 
the latter is the particle size number distribution function, i.e. the number of particles 
per unit volume having radii between $a$ and $a+da$ at height $h$ above ground level. 
More accurately, when vegetation and city structures or outdoor luminaire designs are 
all not efficient enough in blocking light emissions to low elevation angles, an exponential 
term should be kept in Eq. \ref{eq:RS_radiance}, i.e.
\begin{eqnarray}\label{eq:RS_intensity}
    R_{S}(\Theta) &=& \frac{\tilde{R}_{0\xi}}{(\xi \sin \Theta)^{2}} dS \left (
	\frac{\lambda}{2\pi} \right )^{2}
	\nonumber \\
	&& \int_{h=0}^{\xi} exp \left \{[\tau(0)-\tau(h)] \left ( 1-\frac{1}{\cos z'} 
	\right ) \right \}
	\nonumber \\
	&& \int_{a=0}^{\infty} S_{11}(a,z') f(a,h) \sin^{2}(z') ~da~ dh~
\end{eqnarray}
The radiance $R_{S}(\Theta)$ increases steeply when $\Theta$ approaches zero, and we 
evidence in Fig. ~\ref{fig:Rayleigh_vs_Mie} that the aerosol contribution is dominant 
(in comparison with Rayleigh's molecular one) in shaping the radiance patterns for 
almost all emission angles $z'$ and for most typical values of the aerosol asymmetry 
parameter $g$. The latter is commonly used in radiative transfer theories to characterize 
the average cosine of the scattering angle, assuming that the probability of a photon 
to be scattered per unit solid angle around the direction  $z'$ is proportional to the 
phase function $p(z')$. In the numerical demonstration we have varied the asymmetry 
parameter in order to model different aerosol types. Values of $g$ approaching unity 
usually indicate the presence of large particles with a strong forward scattering lobe, 
while $g \approx 0$ corresponds to isotropically scattering media containing particles 
much smaller than the spectral detection range \citep[e.g.][]{MoosmuellerOgren2017}. Most 
typically, $g$ ranges from a few tenths to $\approx$ 0.9. This is why the numerical 
demonstration of aerosol contribution to the satellite radiance is performed 
for $g$=0.3 ... 0.9.

\section{The inverse scattering problem}
\label{sec:inverse} 

\begin{figure}
	\begin{minipage}[t]{0.45\columnwidth}
	\includegraphics[width=1\columnwidth]{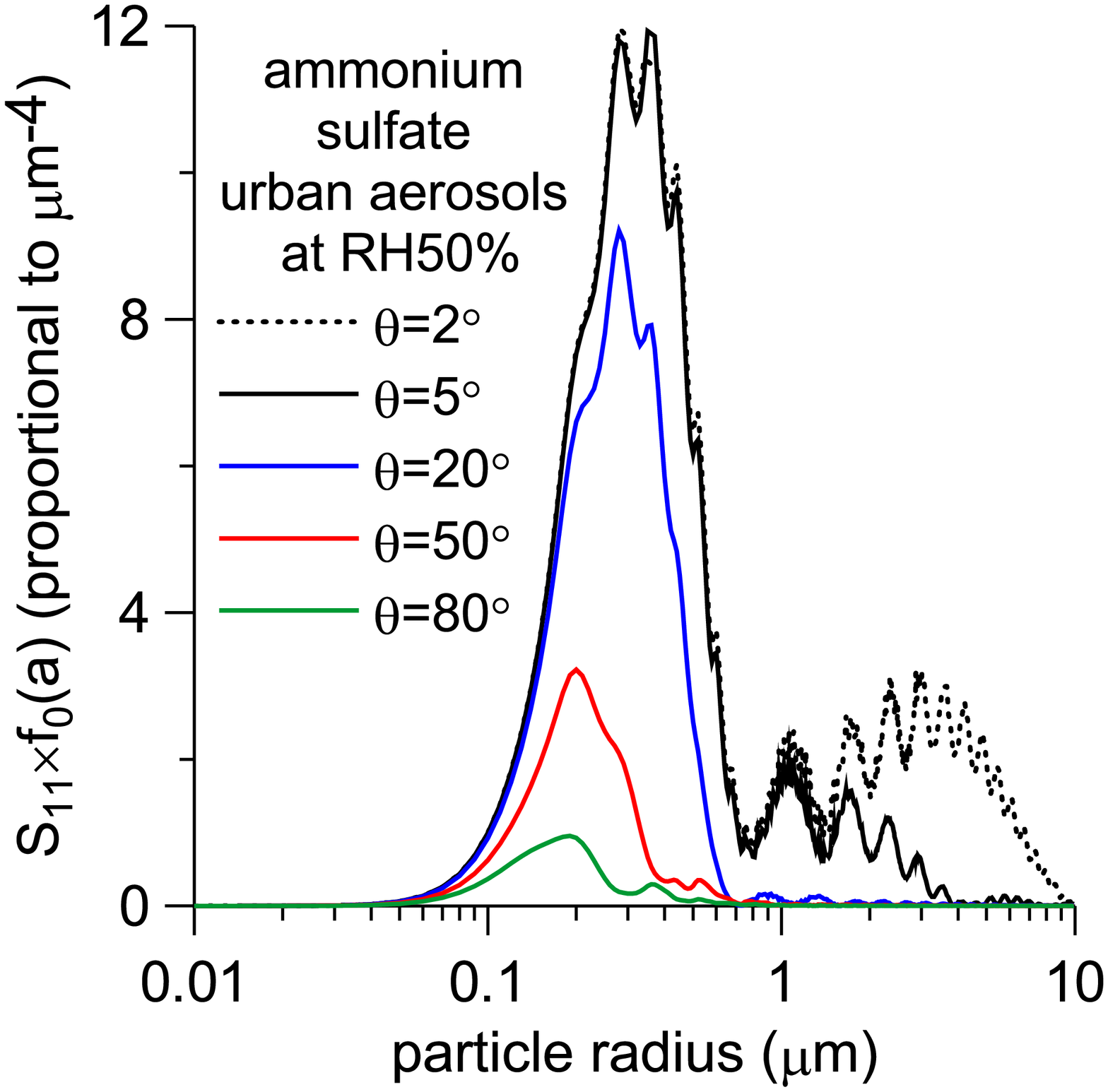}
	\end{minipage}
	\begin{minipage}[t]{0.48\columnwidth} 
	\includegraphics[width=1\columnwidth]{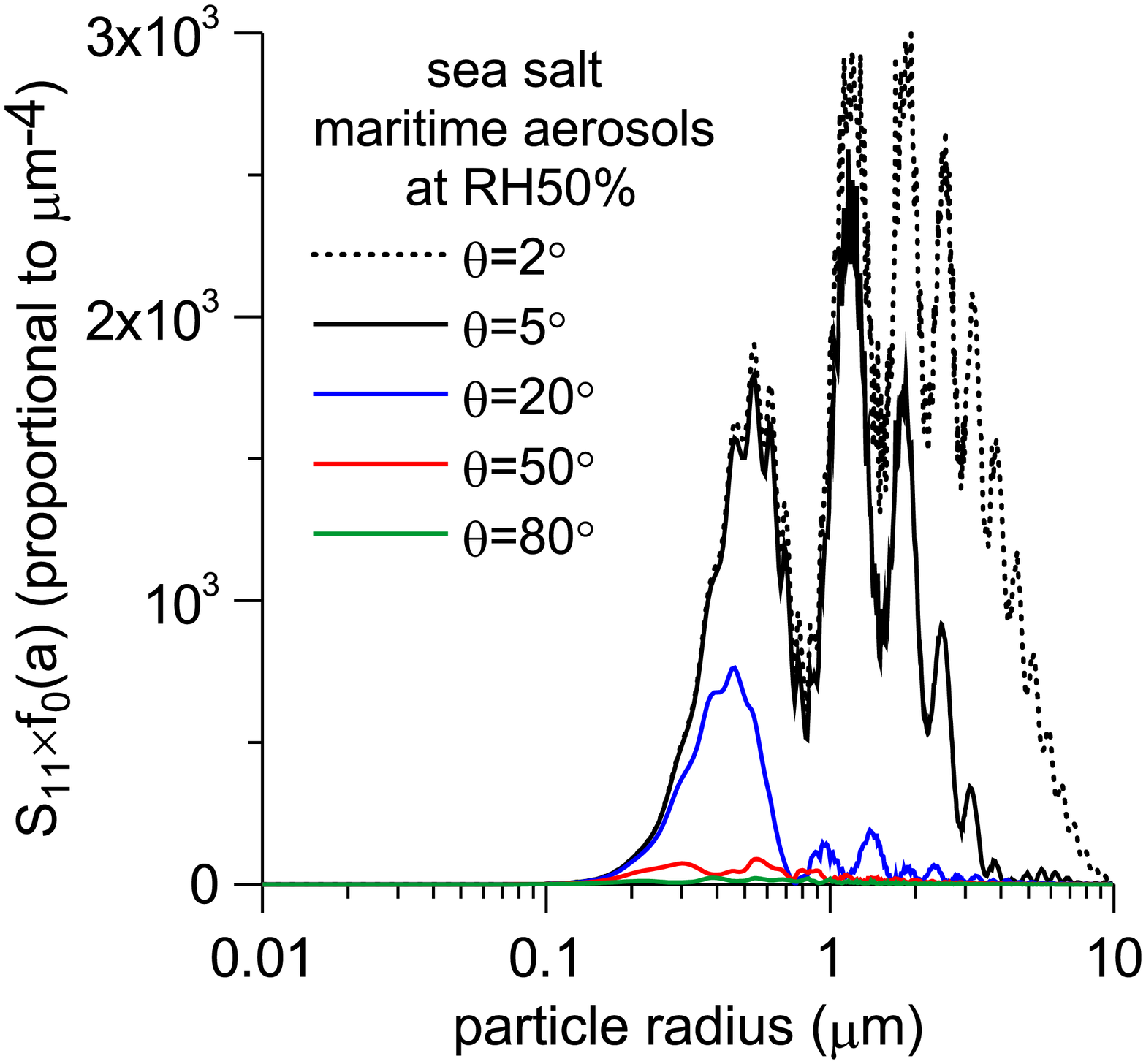}
	\end{minipage} 
    \caption{Relative contribution of different particle radii to the elementary 
	radiance at a set of scattering angles $\theta$ and for relative humidity at
	about 50\%. Left plot (a) is for ammonium sulfate (urban-like) aerosols. 
	Right plot (b) is for sea salt (maritime) particles.}
    \label{fig:sensitivity}
\end{figure}

For a satellite located at near zero zenith angle above the city and $\theta \approx z'$ 
we have $dh \sin^{2} z' = -\xi \sin \Theta dz'$, while the mapping from the space of 
experimental radiance data, $R_{s}(\Theta)$, to the space of particle size distribution 
functions $f_{0}(a)$ is described by the following integral equation
\begin{eqnarray}\label{eq:integralEQ}
    \frac{R_{S}(\Theta_{1})}{\tilde{R}_{0\xi}} &=& \int_{a=0}^{\infty} 
	f_{0}(a)~K_{S11}(\Theta_{1},a)~da~
\end{eqnarray}
where $\Theta_{1}$ is the smallest angular separation between the bright surface 
area of a city and the direction of remote sensing, and $R_{S}(\Theta_{1})$ is 
the radiance scattered towards the satellite from the whole atmospheric column 
along the line of sight ending in point $M$ (see Fig.~\ref{fig:scatt_scheme}b). 
For a vertically stratified 
atmosphere $f(a,h)=f_{0}(a) F(h)$, where $F(h)$ is commonly approximated by an 
exponential function $F(h)=\exp \left \{ -h/H_{A}\right \}$ with $H_{A}$ being 
the aerosol scale height \citep[e.g.][]{WaquetEtAl2007}. In most cases $H_{A}$
ranges from 1 to 3 km. The kernel $K_{S11}(\Theta_{1},a)$ of the integral equation 
has the following form for the satellite remote sensing of a light pollution source
\begin{eqnarray}\label{eq:kernel}
    K_{S11}(\Theta_{1},a) &=& 2 \xi \left ( \frac{\lambda}{2\pi} \right )^{2}
	\int_{\Theta1}^{\Theta1+\Delta \Theta} \epsilon(\Theta)
	\nonumber \\
	&& \int_{z'=0}^{\pi/2} S_{11}(a,z') \exp \left \{ \frac{\xi \sin 
	\Theta}{H_{A} \tan z'} \right \}
	\nonumber \\
	&& dz' d \Theta~
\end{eqnarray}
where $\Delta \Theta = \Theta_{N}-\Theta_{1}$ is the angular span between the nearest 
and farthest bright areas in the city, measured from the point $M$ in the satellite 
images (cfr. Fig.~\ref{fig:scatt_scheme}b). $\epsilon(\Theta)$ is the half-angle 
subtended by each elementary arc of the city emitting area, as seen from the ground 
point $M$, for each value of $\Theta$ (see arcs within dS in Fig.~\ref{fig:scatt_scheme}b). 
The inverse problem consists in finding the solution vector $f_{0}(a)$ that, for a 
given $K_{S11}(\Theta_{1},a)$, produces the best match to the experimentally determined 
data vector $R_{S}(\Theta_{1})/\tilde{R}_{0\xi}$. The retrieval of the particle size 
number distribution function from Eq. \ref{eq:integralEQ} can be difficult because 
the problem is typically ill-posed until additional ({\it a-priori}) information on 
the aerosols can be applied. For instance, the material composition is one of the 
important properties that predetermine the range of applicability of the inverse 
transform. For urban, industrial-like aerosols with refractive index as described in 
\citep[]{KentEtAl1983} we have found that the peak contribution to the scattered signal 
is due to submicrometer-sized particles at almost all scattering angles $\theta$ (see 
Fig.~\ref{fig:sensitivity}a). On the other hand, the contribution from micrometer-sized 
particles to the scattered signal is also important if e.g. humidified sea salt is 
present in the atmosphere (see Fig.~\ref{fig:sensitivity}b). This implies that the 
method is neither sensitive to very small particles with size parameter $x=2 \pi a/\lambda 
\lsim 1$ nor to particles larger than a few micrometers. The traditional solution to 
the Fredholm integral equation (Eq. \ref{eq:integralEQ}) is to implement some suitable 
regularization algorithm, but the problem can be significantly simplified by reducing 
the number of unknowns. For instance, instead of searching for a $N$-element solution 
vector $f_{0}(a_{i})$ ($i=1..N$), an alternative and more efficient approach consists 
in using a parametric analytical model for the particle size number distribution 
function and determining its free parameters by minimizing a cost function, especially 
if the number of these unknown parameters is low. Ideally e.g. a log-normal distribution 
with two shaping parameters (the particle modal radius $r_{M}$ and the size-distribution 
width $\sigma$) may allow for transitioning from a complex inversion problem to a simple 
minimization one.

\section{Processing the VIIRS-DNB nighttime light imagery of Zaragoza city}
\label{sec:inverse} 

\
\begin{figure}
	\begin{minipage}[t]{0.48\columnwidth}
	\includegraphics[width=1\columnwidth]{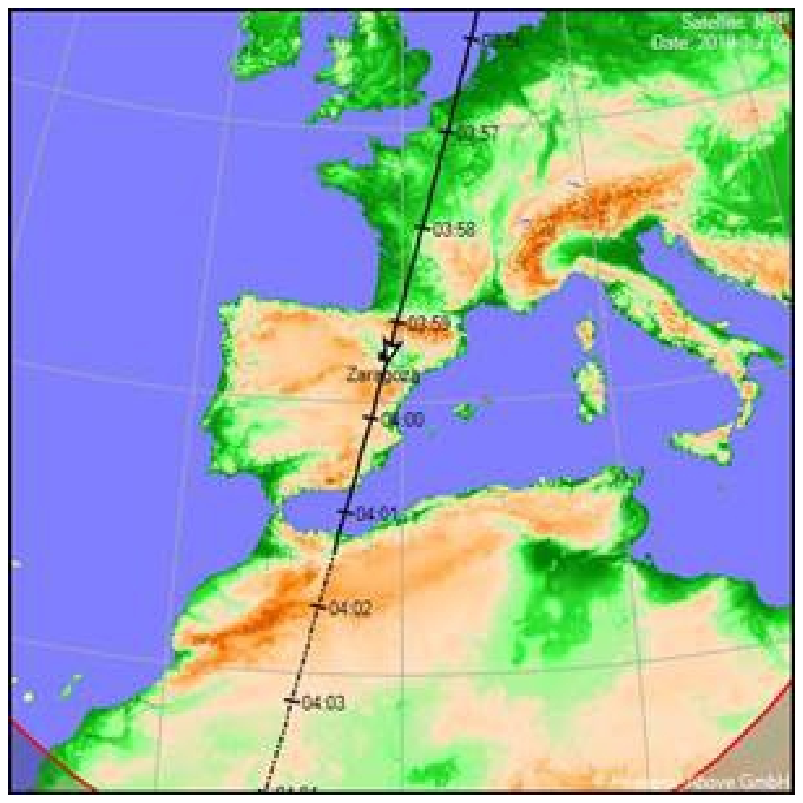}
	\end{minipage}
	\begin{minipage}[t]{0.48\columnwidth} 
	\includegraphics[width=1\columnwidth]{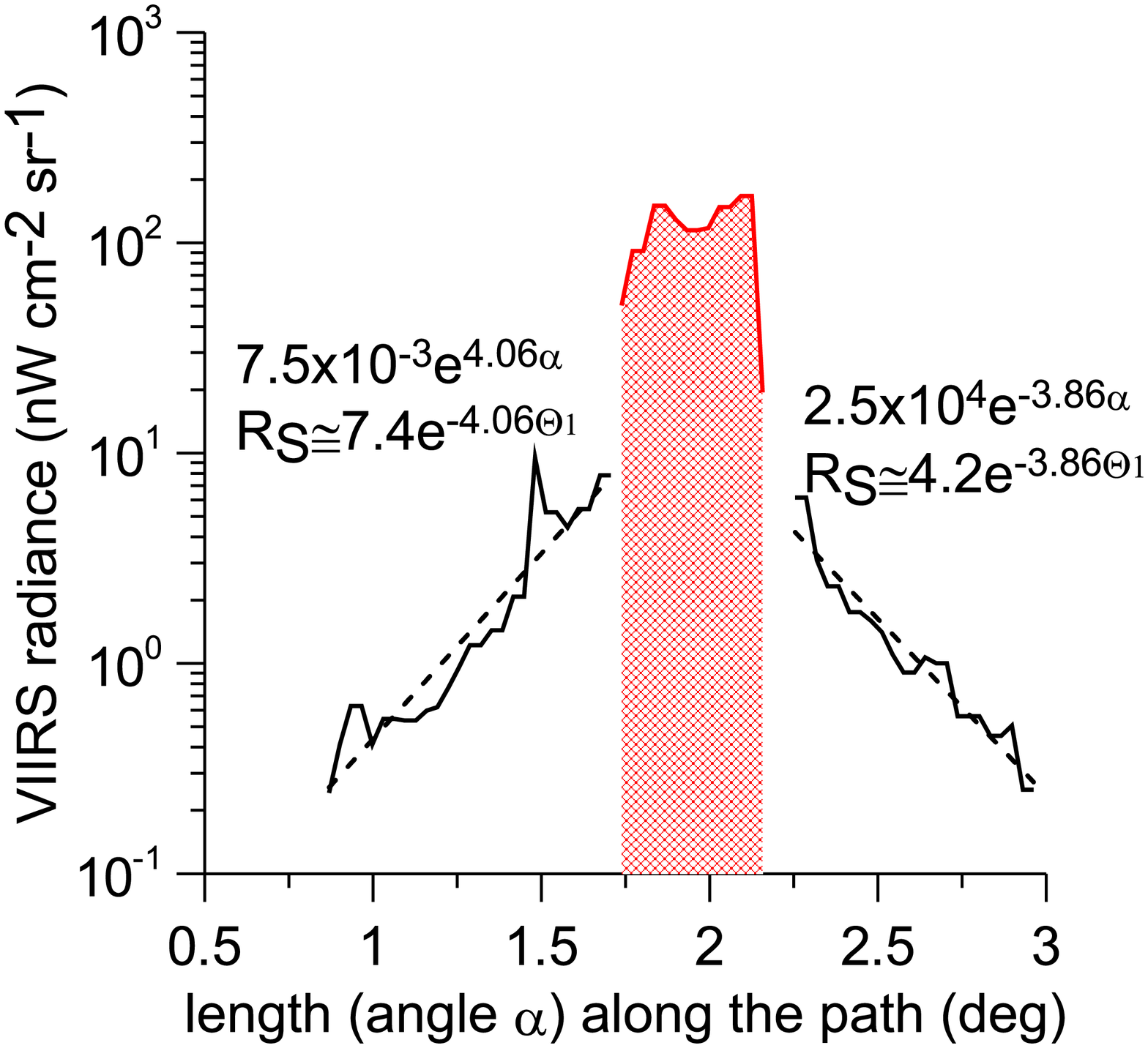}
	\end{minipage} 
    \caption{Left plot (a): projection of the Suomi-NPP satellite's orbit onto the 
	surface of the Earth (ground track), during its pass over Zaragoza (Spain) in 
	the night of July 5, 2019 \citep[]{HeavensAbove2020}. Right plot (b): The 
	radiance as a function of the nadir angle $\alpha$ (measured in degrees, as 
	seen from the satellite). The radiance has been taken at two edges for the 
	first and the last contacts with the artificially lit surface area of the 
	city. Red color is for the inner parts of the city.}
    \label{fig:Zaragoza}
\end{figure}

We present in this section a proof-of-concept of the application of this approach to 
obtain the columnar particle size number distribution function $\tilde{f}_{0}(a)=
\int_{0}^{\infty} f(a,h)~dh=H_{A}~f_{0}(a)$ by inverting Eq. \ref{eq:integralEQ} 
with the kernel given in Eq. \ref{eq:kernel}. We used as input data the radiances 
$R_{S}(\Theta_{1})$ and $\tilde{R}_{0\xi}$ measured by the VIIRS-DNB radiometer 
onboard the Suomi-NPP satellite during its pass over the city of Zaragoza (Spain, 
41.65 N, 0.89 W) in the night of July 5th, 2019 at 03:59 CEST (01:59 UTC). This pass 
took place in a clear and moonless night, with the Moon at 30.14 degrees below the 
horizon, illuminated at 7.2 per cent, and Sun altitude -20.6 degrees, at the time 
of the measurements. That night the satellite passed almost vertically over the the 
city, at an altitude of 89 degrees over the horizon, corresponding to a zenith angle 
$z_{0} = 1^{\circ}$ (see Fig.~\ref{fig:scatt_scheme}a, and ground track in Fig.
~\ref{fig:Zaragoza}a). 

The Suomi-NPP satellite, launched on October 2001, is located in a polar sun-synchronous 
orbit of 826 x 828 km, with an inclination of 98.7$^{\circ}$, recording the Earth radiance 
in a panchromatic band (500-900 nm), with 14-bit quantization and a low light imaging 
detection limit of $5\times10^{-11}~W cm^{-2} sr^{-1}$ \citep[]{CaoAndBai2014, 
ElvidgeEtAl2013, ElvidgeEtAl2017}. The radiance data are available as geotiff files in 
the WGS84 coordinate reference system, with pixel size of 15 arcseconds, measured from 
the center of the Earth  \citep[]{NOAA2019}. This corresponds to a pixel angular size 
of about  $\Delta\Theta$=115 arcseconds in the North-South direction, as seen from the 
satellite location at low nadir angles. The radiance pixel array analyzed in our study 
was taken along a North-South path passing through the urban center, which provides the 
$\tilde{R}_{0\xi}$ values, and encompassing large unpopulated segments of territory at 
its extremes, where the recorded  $R_{S}(\Theta_{1})$ radiance is mostly due to atmospheric 
scattering. Zaragoza has a compact urban structure with well defined city limits, subtending, from the satellite, an angle of order 0.58 degrees. The 
average value $\tilde{R}_{0\xi}$ can be obtained from the radiance recorded within the city 
(see Fig.~\ref{fig:Zaragoza}b, red line). The values of  $R_{S}(\Theta_{1})$ are measured 
from the city border, obtaining in this way two sub-arrays of measurements along the path, 
one before reaching the North border of the city and other starting from the South border 
(Fig.~\ref{fig:Zaragoza}b, full black lines). The dashed black lines in Fig.~
\ref{fig:Zaragoza}b are logarithmic fits of the raw radiance, removing the small local 
peaks associated with some residual sparse sources of light.

Two instrumental data arrays can be immediately computed from these measurements: 
the ratio ${R_{S}(\Theta_{1})}/{\tilde{R}_{0\xi}}$, and the kernel  $K_{S11}(\Theta_{1},a)$ 
(according to Eq. \ref{eq:kernel}). These arrays are the required inputs for performing 
the inversion of Eq. \ref{eq:integralEQ} in order to obtain the estimate $\tilde{f}_{0}(a)$ . 
This inversion can be carried out using different approaches. One of them is to perform a 
conventional linear estimation with an appropriate Tikhonov's regularization parameter. 
Another useful procedure consists on the choice of a suitable analytic form for 
$\tilde{f}_{0}(a)$, based on the available {\it a-priori} knowledge about the physical 
properties of the phenomenon under study, and determining the unknown parameters of this 
trial function by using conventional minimization routines. In this work we resorted to the 
latter option, by assuming that the particle size number distribution function can be well 
described by the linear combination of a set of i=1...$M$ log-normal distributions in the form
\begin{eqnarray}\label{eq:logNormal}
   \tilde{f}_{0}(a) = \sum_{i=1}^{M} \frac{N_{i}}{\sigma_{i} a} 
\exp \left \{ -\frac{1}{2} \frac{\left [ \log_{10}(a)-\log_{10}(a_{i}) 
\right ]}{\sigma_{i}^{2}} \right \}
\end{eqnarray}
where each ${i}$-th distribution is characterized by a small set of parameters ($a_{i}, 
\sigma_{i}, N_{i}$), informing about its position and width on the particle size space, 
and its relative weight in the linear combination, respectively. Ideally a low number of 
elementary distributions ($M$=1 or 2) should be enough for building up a good approximation 
to $\tilde{f}_{0}(a)$. Some additional {\it a-priori} knowledge may be used to improve the 
estimation. Zaragoza is located in the mid of a semi-arid zone, so that dust-like particles 
make an important part of the total aerosol content. The characteristic refractive index 
of dust-like particles, 1.52+0.025$i$, can be potentially influenced by humidity at night, 
so we used a reduced index of 1.4+0.002$i$. With these assumptions we minimized the quadratic 
differences between the measured radiance ratios, ${R_{S}(\Theta_{1})}/{\tilde{R}_{0\xi}}$, 
and those expected from Eq. \ref{eq:logNormal} (substituted into Eqs. \ref{eq:integralEQ} 
and \ref{eq:kernel}) with respect to the unknown parameters $a_{i}, \sigma_{i}, N_{i}$ ($M$=2), 
avoiding too narrow and too wide widths ($\sigma_{i}$), as well as to small and too large 
modal radii $a_{i}$. The computations were made for $\lambda$ = 650 nm, the weighted average 
wavelength of the VIIRS-DNB spectral sensitivity band.

\begin{figure}
	\begin{minipage}[t]{0.49\columnwidth}
	\includegraphics[width=1\columnwidth]{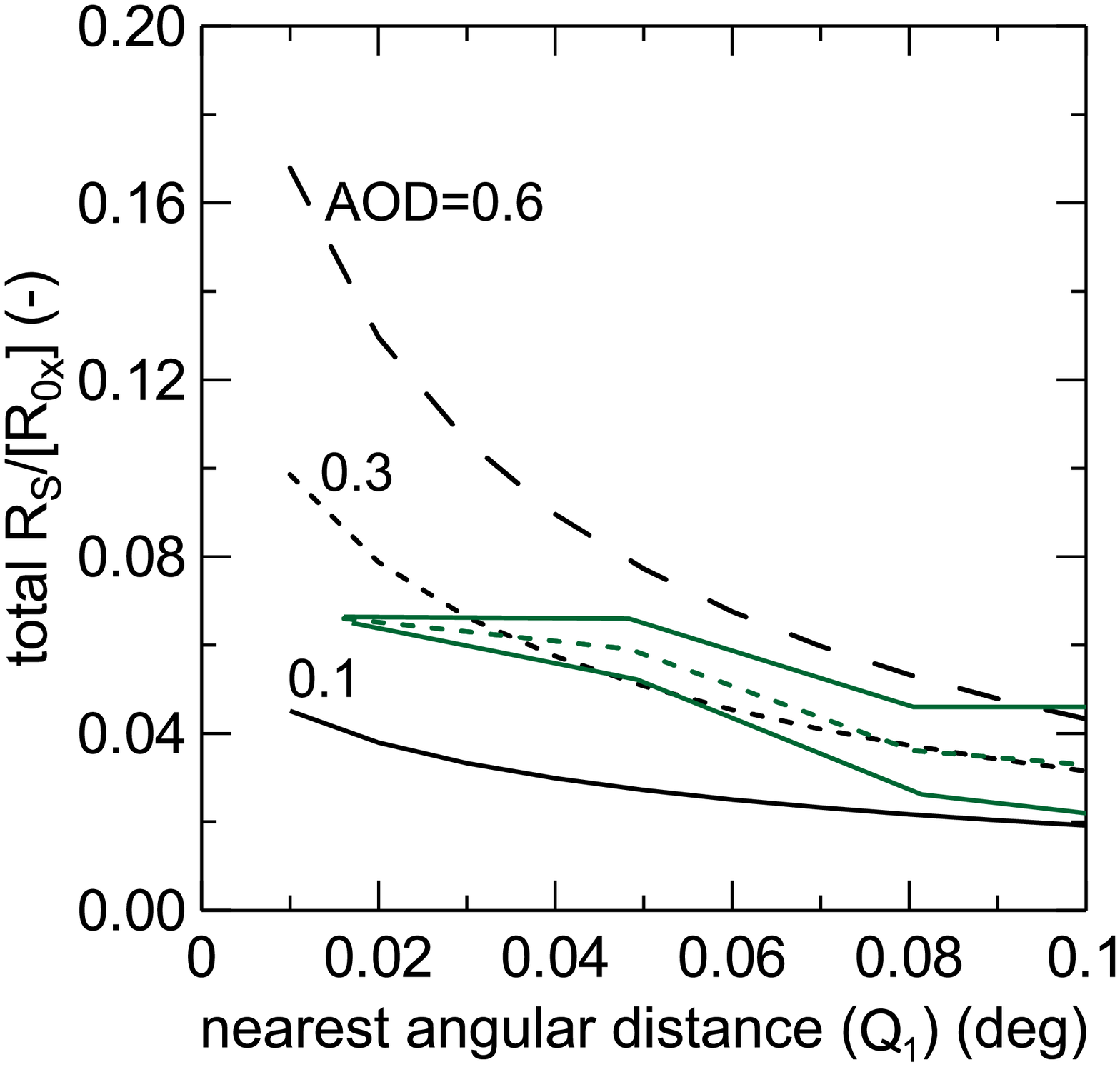}
	\end{minipage}
	\begin{minipage}[t]{0.47\columnwidth} 
	\includegraphics[width=1\columnwidth]{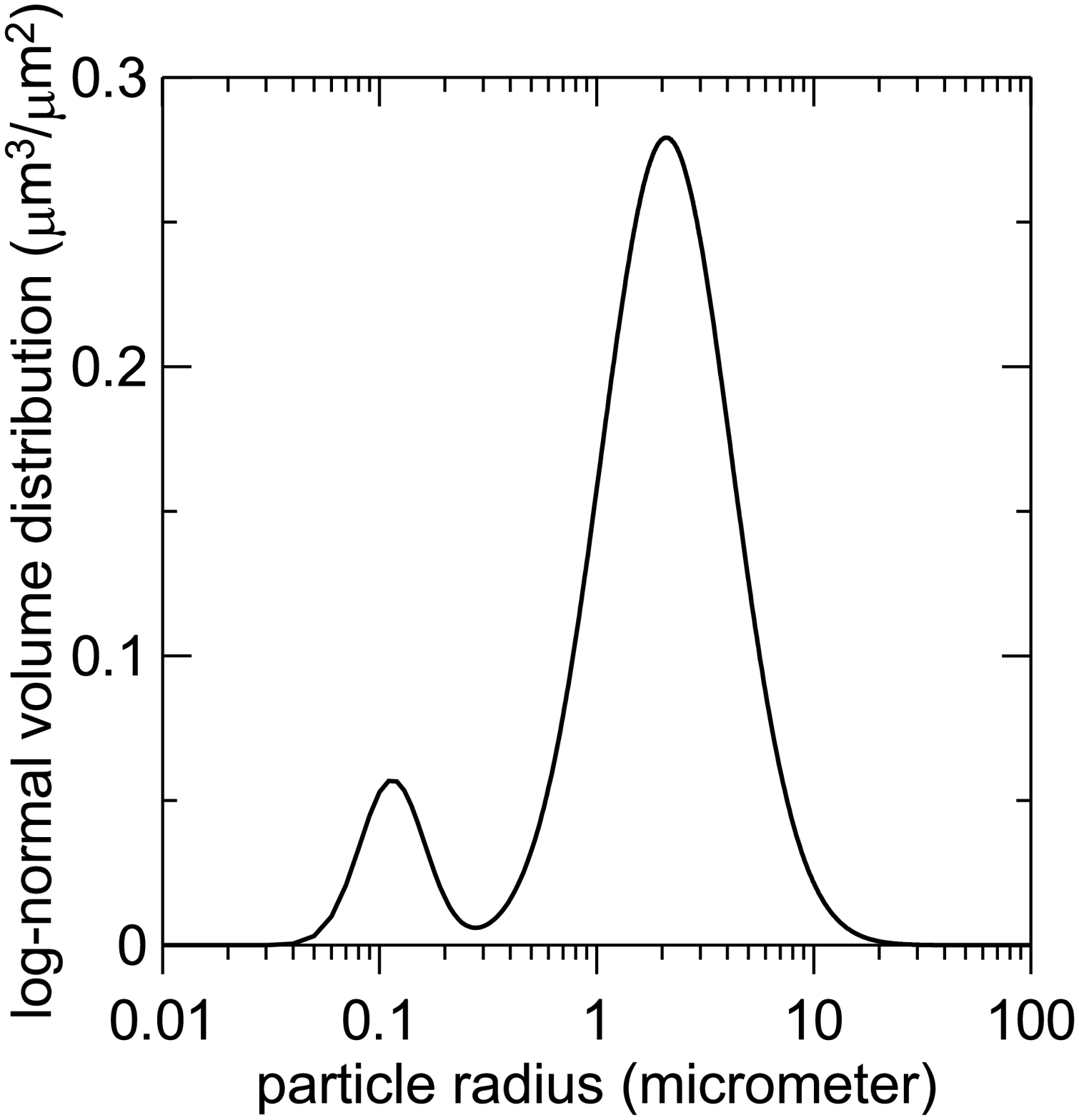}
	\end{minipage} 
    \caption{Left plot (a): Modelled (black lines) versus measured (green lines)
	ratios $R_{S}(\Theta_{1})/\tilde{R}_{0\xi}$ for different aerosol optical depths
	at $\lambda$=650nm. The solid green lines correspond to the two experimentally determined
	data sets, while the dashed line is the weighted average of both. Right plot (b): The 
	particle size distribution function determined by minimization of the differences between 
	the model and the measurements and presented the same way as in AERONET database, i.e.
	$dV/d\ln a=a~dV/da=(4/3)\pi a^{4} \tilde{f}_{0}(a)$, where $dV/da$, where $dV/da$ is the 
	particle size volume distribution function and $\tilde{f}_{0}(a)=\int_{0}^{\infty} 
	f(a,h)~dh=H_{A}~f_{0}(a)$ is the columnar particle size number distribution.}
    \label{fig:Zaragoza_results}
\end{figure}

Incorporating all these constraints we obtain $N_{1}/N_{2} \approx 125$, $a_{1} \approx$ 
0.08 $\mu m$, $\sigma_{1} \approx$ 0.15, $a_{2} \approx$ 0.5 $\mu m$, and $\sigma_{2} \approx$ 
0.3. The resulting particle distribution function is shown in form of the volume density 
function (Fig.~\ref{fig:Zaragoza_results}b) to be consistent with presentation form of 
the AERONET data products. Fig. \ref{fig:Zaragoza_results}a shows that the best match 
of the theoretical model (Eq.~\ref{eq:RS_radiance}) to the experimental data occurs for AOD 
$\approx$ 0.3. The $R_{S}(\Theta_{1})/\tilde{R}_{0\xi}$ ratio as a function of $\Theta_{1}$ 
becomes steeper for larger AODs, and more flat for lower AODs. In addition, the absolute values 
of $R_{S}(\Theta_{1})/\tilde{R}_{0\xi}$ increase with increasing AOD, so one can easily 
estimate the optimum AOD for the time of measurement. These results are fairly similar to 
those reported by the AERONET station located in Zaragoza (41.63 N, 0.88 W, height=250 m) a 
few hours after the Suomi-NPP pass \citep[]{NASA2020}: the recorded AOD at ~05:45 UTC for 
$\lambda$=675 nm was 0.3, and the retrieved particle size volume distribution function clearly 
shows the same bimodal structure predicted in Fig. \ref{fig:Zaragoza_results}b, a feature 
of the aerosol distribution that was maintained along that day. The two peaks of that function 
at 06:07 UTC, first available data for this day, were centered in aerosol particle radii 
slightly above 0.1 and 2 $\mu m$, with maximum values close to 0.035 and 0.13 $\mu m^{3}\mu m^{-2}$, 
respectively. The peak locations are similar, and the peak values are of the same order of 
magnitude but slightly lower than those deduced from the VIIRS-DNB images. Some differences 
could be expected, since the VIIRS-DNB measurements were obtained at 01:59 UTC, a few hours 
before dawn.

Let us point out that the actual particle volume distribution function is not expected 
to be as smooth as that displayed in Fig.~\ref{fig:Zaragoza_results}b. The smoothness of 
this estimate comes from the fact that it has been obtained using a linear combination 
of smooth log-normal elementary distributions, as defined in Eq. \ref{eq:logNormal}. A 
more precise estimate of this function could be obtained by inverting Eq.~\ref{eq:integralEQ} 
using a suitable, but numerically more demanding, regularization procedure.  

\section{Conclusions}

We demonstrate in this work that the diffuse atmospheric scattered radiance distribution 
around urban nuclei detected by on-orbit radiometers at night-time contains useful and 
retrievable information about some key aerosol properties, like the columnar number size 
distribution function. The method is self-calibrating and does not require the knowledge 
of the absolute city emissions. This opens the way for more comprehensive studies of the 
aerosol dynamics at night, improving our knowledge of their characteristics beyond the 
traditional aerosol optical depth. The results obtained with this approach applied to 
the radiance data obtained by the VIIRS-DNB radiometer in a particular pass over the 
city of Zaragoza are presented as a proof-of concept of its feasibility and performance. The DNB band is a panchromatic one, extending from the visible to the NIR. It can be anticipated that the use of multiband radiometry (e.g. RGB calibrated images of night lights acquired from the International Space Station) may provide additional information for a more detailed aerosol characterization.

\section*{Acknowledgements}

This work was supported by the Slovak Research and Development Agency under contract 
No: APVV-18-0014. Computational work was supported by the Slovak National Grant Agency 
VEGA (grant No. 2/0010/20). The authors thank Emilio Rodr\'{\i}guez Fern\'{a}ndez from 
Universidade de Santiago de Compostela for his help in locating suitable Suomi-NPP 
pasess over cities, Chris Peat from Heavens-Above by granting permission for reproducing 
the Suomi-NPP ground track image, and Juan Ram\'{o}n Moreta Gonz\'{a}lez for kindly 
providing the data of the AERONET station in Zaragoza. Thanks are also due to the Reviewer for useful suggestions and comments.





\begin{thebibliography}{99}
\bibitem[\protect\citeauthoryear{Cao and Bai}{2014}]{CaoAndBai2014}
Cao C., Bai Y., 2014, Remote Sens. 6, 11915, doi:10.3390/rs61211915
\bibitem[\protect\citeauthoryear{Choo and Jeong}{2016}]{ChooJeong2016}
Choo G. H., Jeong M. J., 2016, Korean Journal of Remote Sensing, 32, 73
\bibitem[\protect\citeauthoryear{Elvidge et al.}{2013}]{ElvidgeEtAl2013}
Elvidge C. D., Baugh K., Zhizhin M., Hsu F. C., 2013, Proceedings of the Asia-Pacific 
Advanced Network 2013, 35, 62, doi:10.7125/APAN.35.7
\bibitem[\protect\citeauthoryear{Elvidge et al.}{2017}]{ElvidgeEtAl2017}
Elvidge C. D. , Baugh K., Zhizhin M., Hsu F. C., Ghosh T., 2017, International Journal 
of Remote Sensing 38, 5860. DOI: 10.1080/01431161.2017.1342050
\bibitem[\protect\citeauthoryear{Johnson et al.}{2013}]{JohnsonEtAl2012}
Johnson R. S., Zhang J., Hyer E. J., Miller S. D., Reid J. S., 2013, Atmos. Meas. 
Tech. 6, 1245, https://doi.org/10.5194/amt-6-1245-2013, 2013
\bibitem[\protect\citeauthoryear{Kent et al.}{1983}]{KentEtAl1983}
Kent G. S., Yue G. Y., Farrukh U. O., Deepak A., 1983, Appl. Opt. 22, 1655
\bibitem[\protect\citeauthoryear{Kokhanovsky}{1998}]{Kokhanovsky1998}
Kokhanovsky A.~A., 1998, Journal of the Atmospheric Sciences. 55, 314
\bibitem[\protect\citeauthoryear{Koll\'{a}th and Kr\'{a}nicz}{2014}]{KollathEtAl2014}
Koll\'{a}th Z., Kr\'{a}nicz B., 2014, Journal of Quantitative Spectroscopy 
and Radiative Transfer, 139, 27
\bibitem[\protect\citeauthoryear{McHardy et al.}{2015}]{McHardyEtAl2015}
McHardy T. M., Zhang J., Reid J. S., Miller S. D., Hyer E. J., Kuehn, R. E., 
2015, Meas. Tech., 8, 4773. https://doi.org/10.5194/amt-8-4773-2015 
\bibitem[\protect\citeauthoryear{Moosm\"{u}ller and Ogren}{2017}]{MoosmuellerOgren2017}
Moosm\"{u}ller H, Ogren J. A., 2017, Atmosphere 8, 133
\bibitem[\protect\citeauthoryear{NASA}{2020}]{NASA2020}
NASA, 2020, Aerosol Robotic Network (AERONET), \url{https://aeronet.gsfc.nasa.gov/cgi-bin/draw_map_display_aod_v3} 
(Last accessed, March 13, 2020)
\bibitem[\protect\citeauthoryear{NOAA}{2019}]{NOAA2019}
NOAA, Earth Observation Group, 2019, VIIRS Daily Mosaic, 
\url{https://ngdc.noaa.gov/eog/viirs/download_ut_mos.html} (Last accessed, 20 July 2019)
\bibitem[\protect\citeauthoryear{Peat}{2020}]{HeavensAbove2020}
Peat C., 2020, Heavens-Above,  \url{https://www.heavens-above.com/} (Last accessed, 15 March 2020)
\bibitem[\protect\citeauthoryear{S\'{a}nchez de Miguel et al.}{2019}]{SanchezEtAl2019}
S\'{a}nchez de Miguel A., Kyba C. C. M., Zamorano J., Gallego J., Gaston K.J., 
2019, arXiv:1908.05482 [astro-ph.IM], https://arxiv.org/abs/1908.05482
\bibitem[\protect\citeauthoryear{Wang et al.}{2016}]{WangEtAl2016}
Wang J., Aegerter C., Xu X., Szykman J. J., 2016,  Atmos. Environ., 124, 55–63, 2016.
\bibitem[\protect\citeauthoryear{Waquet et al.}{2007}]{WaquetEtAl2007}
Waquet F., Goloub, P., Deuz\'{e} J. L., L\'{e}on J. F., Auriol F., Verwaerde C., Balois
J. Y., Francois P., 2007, J. Geophys. Res. 112, D11214
\bibitem[\protect\citeauthoryear{Zhang et al}{2008}]{ZhangEtAl2008}
Zhang J., Reid J. S., Turk J., Miller, S., 2008, Int. J. Remote Sens., 29, 4599
\bibitem[\protect\citeauthoryear{Zhang et al}{2019}]{ZhangEtAl2019}
Zhang J., Jaker S. L., Reid J. S., Miller S. D., Solbrig J., Toth T. D., 2019, Atmos. Meas. 
Tech., 12, 3209. https://doi.org/10.5194/amt-12-3209-2019
\end{thebibliography}




\bsp	
\label{lastpage}
\end{document}